# Magnetic-Field Control of Emergent Order in a 3D Dipolar Pyramid Artificial Spin Ice

Luca Berchialla,[1,2,*] Gavin M. Macauley,[1,2,†] Flavien Museur,[1,2] Anja Weber[1,2,‡] and Laura J. Heyderman[1,2,*]

[1] Laboratory for Mesoscopic Systems, Department of Materials, ETH Zurich, 8093 Zurich, Switzerland

[2] PSI Center for Neutron and Muon Sciences, 5232 Villigen PSI, Switzerland

[†] Present address: Department of Physics, Princeton University, Princeton, NJ 08540 USA

[‡] Present address: PSI Center for Life Sciences, 5232 Villigen PSI, Switzerland

**ABSTRACT**. We realize a three-dimensional artificial spin ice of disconnected nanomagnets interacting solely via dipolar coupling, patterned on square pyramids. This Pyramid artificial spin ice, with both tilted and in-plane nanomagnets, supports tunable states. Monte Carlo simulations reveal a rich phase diagram and an emergent square ice of vertex-level effective spins. Tailored demagnetization protocols and magnetic force microscopy allow experimental access to low-energy states, establishing a platform for exploring three-dimensional artificial spin ices.

## I. INTRODUCTION.

Recent breakthroughs in nanofabrication have enabled the creation of three-dimensional (3D) magnetic structures and devices at the mesoscale, going beyond the limits of 2D thin film magnetism [1,2]. In parallel, magnetic characterization techniques, both established and new, have begun to reveal the rich and surprising behavior that arises in 3D environments [3–6] .

Among 3D magnetic systems, artificial spin ices (ASIs) [7], which are precisely engineered arrays of nanomagnets mimicking the behavior of frustrated spin systems [8], stand out as a powerful means to investigate emergent magnetic phenomena. Harnessing the third dimension for ASIs has always been desirable since it would enable access to exotic magnetic states with new functionalities, which are potentially useful for applications such as computation [7]. The possibilities to make and realize 3D ASIs have been explored since the early days of the field, first theoretically [9–11], and later experimentally, with multilayered ASIs, in which nanomagnets are arranged within multiple planes above and parallel to the surface of a flat substrate [8,12–16]. However, fabricating and imaging a truly 3D arrangement of single-domain nanomagnets, with the nanomagnets not only arranged on different planes but also oriented at an angle to the substrate, is a significant challenge. To date, the only examples of 3D ASIs are based on interconnected 3D nanowire lattices [17–23]. In such nanowire lattices, each link of the lattice is single domain, much like the single-domain Ising-like nanomagnets common to two-dimensional (2D) ASIs. However, when these nanomagnets are physically joined at a vertex in an interconnected lattice, whether in 2D or 3D, the interaction between the nanomagnets is primarily exchange-driven, with a domain wall frequently found at a vertex [19,20,24,25]. This is in stark contrast to 2D or multilayered ASIs with separated nanomagnets, where the interactions are purely dipolar.

Magnetic imaging of 3D nanostructures is challenging, whether using synchrotron x-ray methods, requiring specialist knowledge and long acquisition times, or laboratory-based techniques such as magnetic force microscopy (MFM) [19,20,22,26]. MFM relies on the precise tracking of the surface topography with the magnetic tip and, because of this, imaging of magnetic configurations in 3D ASI is significantly hindered by the presence of deep valleys between nanomagnets.

Here, we present the first realization of a 3D ASI with physically-separated nanomagnets that interact solely via the magnetic dipolar interaction. This system is based on a novel lattice design with the nanomagnets arranged on an array of square-based pyramids, which we refer to as a Pyramid ASI, and displays a rich phase diagram on varying the orientation of the lateral faces of the pyramids and the strength of the external out-of-plane magnetic field. Thanks to its continuous topography we can unambiguously determine the magnetic configurations using MFM, and we show that, with tailored demagnetization protocols, we can access selected low-energy states. These results highlight the complexity and possibilities offered by this new 3D ASI platform.

## II. PYRAMID ARTIFICIAL SPIN ICE DESIGN

To realize our 3D Pyramid ASI composed of dipolar-coupled, physically-separated nanomagnets, we pattern nanomagnets onto an array of square-based pyramids [Fig. 1(a)]. Half of the nanomagnets [shown in blue in Figs. 1(a-c)] are patterned onto the lateral, sloped faces of the pyramids, while the other half of the nanomagnets [shown in yellow in Figs. 1(a-c)] are patterned on the in-plane regions of the substrate between pyramids. We refer to these two sets of nanomagnets as tilted and in-plane nanomagnets, respectively. This design results in a square lattice arrangement of nanomagnets when viewed from above [Fig. 1(b)].

The pyramids are fabricated by reactive ion etching (RIE) of a silicon (100) substrate (see End Matter, Appendix A). In patterning the Pyramid ASI, the distance between the nanomagnet ends and the vertices is kept the same for all vertices, which means that tilted nanomagnets are longer than their flat counterparts.

Based on which set of nanomagnets meet at a vertex, three distinct vertex topologies can be identified: flat vertices, formed by four in-plane nanomagnets; pyramid vertices, formed by four tilted nanomagnets meeting together at the apex of each pyramid; and mixed vertices, where two tilted nanomagnets and two in-plane nanomagnets meet in the valley between two pyramids. As in the artificial square ice [27], we can define four vertex types, Type I to Type IV, based on how the magnetic moments (or magnetization) associated with the four nanomagnets – so-called macrospins, are arranged at a vertex [Fig. 1(d)].

The macrospin configurations can be unambiguously determined using MFM, which is facilitated in the Pyramid ASI by the fact that the magnetic tip scans over a continuous sample surface and the orientation of the macrospin in discrete nanomagnets is easily determined from the distinct magnetic contrast. This is in contrast to the more complex imaging conditions in nanowire lattices [19,20,22]. For the in-plane nanomagnets, the macrospin points from the bright to the dark end, as in 2D ASIs [27]. For tilted nanomagnets, the magnetic contrast of the nanomagnet ends near the apex of the pyramid, rather than those in the valleys, provides a more reliable signature with the strength of the magnetic signal from all nanomagnets at the apex being the same. In Fig. 1(d), we show representative examples of the MFM contrast at the pyramid apex for each vertex type. By averaging the magnetic signal within the red boxes indicated in Fig.1(d), we can determine whether the macrospin of each nanomagnet points towards (away) from the vertex, reflected by the bright yellow (dark blue) contrast close to the vertex.

*Contact author: l.berchialla@gmail.com
*Contact author: laura.heyderman@psi.ch

An important feature of our system is that an out-of-plane magnetic field will only switch the magnetization in the tilted nanomagnets, leaving the in-plane nanomagnets unaffected. As we will show later, the application of an out-of-plane field results in the emergence of a coarse-grained artificial square ice, where the net moments of Type III vertices act as effective macrospins, thus doubling the periodicity of the lattice.

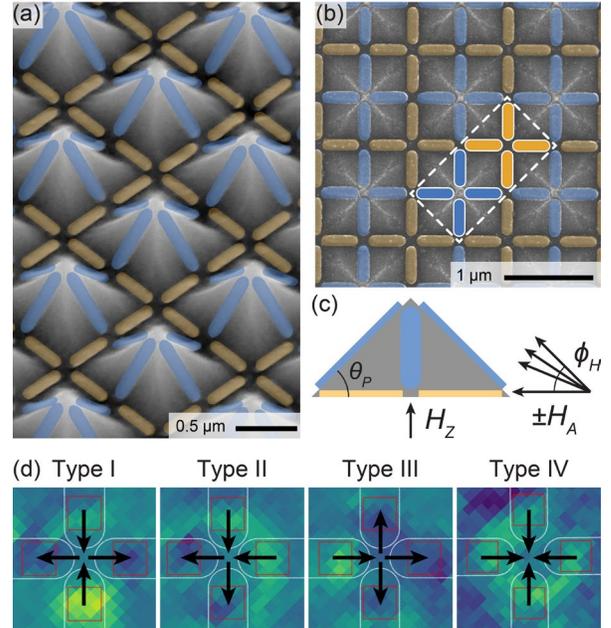

FIG 1. Design of the Pyramid ASI. Scanning electron micrographs of a Pyramid ASI viewed from (a) an oblique angle and (b) from above. Tilted nanomagnets on the lateral faces of the pyramids and in-plane nanomagnets arranged between the pyramids are colored in blue and yellow, respectively. The unit cell encompassing 8 nanomagnets is highlighted with a dashed white box in panel (b). (c) Schematic side view of a single pyramid, showing the angle between the square base and triangular face $\theta_P$ and the angle $\Phi_H$ that defines the orientation of the applied magnetic field $H_A$ during a demagnetization protocol. (d) Examples of magnetic phase contrast obtained with MFM of the pyramid vertices for the different vertex types. The edges of nanomagnets are delineated in white, while the areas used to calculate the average signal are in enclosed in red boxes. The black arrows show the macrospin orientation.

## III. DETERMINING THE PHASE DIAGRAM

We use Monte Carlo simulations based on a point-dipolar Hamiltonian (see Supplemental Material [28], Section S1) to construct the phase diagram of the Pyramid ASI as a function of the pyramid face angle $\theta_P$ and the strength of an out-of-plane magnetic field

$H_Z$ [geometry given in Fig.1(c)]. We find that the phase diagram, illustrated in Fig. 2, features five different low-energy regimes in different regions of the $(\theta_P, H_Z)$-plane. The out-of-plane magnetic field $H_Z$ only affects the magnetization of the tilted nanomagnets, while altering the pyramid base angle $\theta_P$ modifies the lowest-energy configuration of mixed vertices. In particular, for mixed vertices, the lowest-energy configuration is Type I for $\theta_P < 50°$, and Type II for $\theta_P > 50°$, unless $H_Z$ is high enough to polarize all tilted nanomagnets. In addition, for $\theta_P > 0°$, the eight possible Type III vertex configurations occupy two energy levels (see Supplemental Material [28], Section S2). We will not differentiate between them here since this is not essential for the understanding of the behavior of the Pyramid ASI.

We now describe the magnetic phases for low, medium and high $H_Z$. The change in the lowest-energy vertex configuration for mixed vertices on increasing $\theta_P$ means that, in general, we can distinguish between two regimes: one for $0° < \theta_P < 50°$ and one for $\theta_P > 50°$.

*Low-energy states for low $H_Z$*—For $\theta_P < 50°$, the Pyramid ASI has a doubly-degenerate ground state containing Type I vertices equivalent to that found in the 2D artificial square ice [Fig. 2(i)]. For $\theta_P > 50°$, the mixed vertices have a Type II configuration, with their net moments aligning to form chains of head-to-tail effective moments leading to a two-fold degenerate ground state [Fig. 1(ii)]. Here, the ferromagnetically-ordered net moments in adjacent chains are oriented antiparallel to one another, giving a type of collinear antiferromagnetic order [29,30].

*Low-energy states for medium $H_Z$*—Increasing the external magnetic field $H_Z$ seeds Type III configurations on the pyramid vertices. This results in Type III vertices on at least two of the adjacent mixed vertices. Flat vertices, where all of the nanomagnets are in-plane, always adopt Type I configurations. By tuning $\theta_P$, the configuration of the remaining two mixed vertices can be tuned to give either Type I or Type II configurations. For $\theta_P < 50°$, the net moments of Type III vertex configurations on pyramid and mixed vertices form chains of effective magnetic moments in alternating directions, as in a collinear antiferromagnet [Fig. 2(iii)]. For $\theta_P > 50°$, the net moments of Type II and Type III vertex configurations align in a complex antiferromagnetic state [Fig. 2(iv)]. Both of these low-energy states are two-fold degenerate for a given direction of $H_Z$.

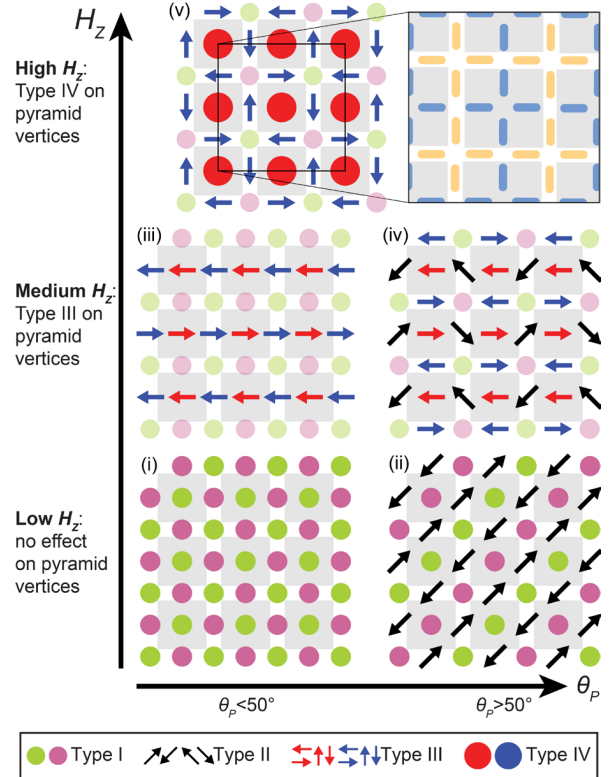

FIG 2. Phase diagram of the Pyramid ASI as a function of the pyramid face angle $\theta_P$ and out-of-plane magnetic field $H_Z$. All configurations shown are obtained from Monte Carlo simulations. For sub-panels iii, iv and v, Type I vertex configurations are displayed in a lighter color to facilitate the visualization of the magnetic state associated with the remaining vertices. Sub-panel v is for all $\theta_P > 0°$. The pyramids appear in the schematics as light gray squares.

*Low-energy states for high $H_Z$*—For sufficiently strong $H_Z$ and for $\theta_P > 0°$, the external field magnetizes all tilted nanomagnets, so that pyramid vertices are all in a Type IV configuration. This results in the vertex configurations of all mixed vertices to be Type III. Remarkably, the net magnetic moments of the mixed vertices form a secondary lattice of effective macrospins that arrange themselves into head-to-tail loops, which is analogous to the ground state of a square ASI. This emergent, coarse-grained square ice has twice the periodicity of the underlying nanomagnet lattice, where the effective spins are vertex-level net moments rather than those associated with individual nanomagnets. In the next section we will describe the thermal properties of this emergent square ice and show how it can be accessed experimentally through a tailored demagnetization protocol.

*Contact author: l.berchialla@gmail.com
*Contact author: laura.heyderman@psi.ch

## IV. EXPERIMENTALLY OBTAINING THE LOW ENERGY STATES

Experimentally, we realized the Pyramid ASI with a fixed $\theta_P \sim 45°$. Keeping the same distance between the nanomagnet ends at all vertices results in tilted nanomagnets being longer than in-plane nanomagnets by a factor of $\sqrt{2}$, increasing their switching field [31] by ~5% (see Supplemental Material [28], Section S3). The blocking temperature of an isolated nanomagnet depends on its volume. Because of this, reaching low-energy states through a thermally-activated process would be challenging, since it is likely that the tilted and in-plane nanomagnets will freeze on different temperature scales. To circumvent this, we instead employ a demagnetization protocol applying an oscillating bipolar magnetic field $H_A$, applied at four angles $\Phi_H = \{0°, 22.5°, 33.75°, 45°\}$ schematically shown in Fig. 1(c) while rotating the sample around an axis perpendicular to the substrate plane (see End Matter, Appendix B for details). By adjusting $\Phi_H$, we can tune the component of $H_A$ acting along the long axis of in-plane and tilted nanomagnets so that both sets of nanomagnets approach their coercive fields simultaneously.

For each $\Phi_H$, we image the resulting magnetic configuration of 15 nominally identical Pyramid ASIs using MFM (see Supplemental Material [28], Section S4). We then extract the vertex populations from each configuration and compute their averages to evaluate the effectiveness of the demagnetization protocol [Fig. 3(a)]. While a perfect demagnetization would bring the system into the zero-field ground state for $\theta_P < 50°$ [Fig. 2(i)], which consists entirely of alternating Type I vertices, our protocol only partially approaches this state, reaching a maximum Type I population of 37% at $\Phi_H = 45°$ [blue data in Fig. 3(a)]. A fully ordered ground state is difficult to achieve because of the 3D nature of the Pyramid ASI. It is surprising that, experimentally, the optimal field angle for approaching this low-energy state is $\Phi_H = 45°$. Ideally, the optimal field angle would be $\Phi_H = 22.5°$, since this would give equal field components along the in-plane nanomagnets and the tilted nanomagnets orientated at $\theta_P \sim 45°$. Even considering the slight curvature of tilted nanomagnets, and their higher coercivity compared with that of the in-plane nanomagnets (see Supplemental Material [28], Sections S3 and S5), we would expect that the optimal field angle to achieve the ground state is $\Phi_H \sim 30°$. However, for $\Phi_H = 0°$ or 22.5° the demagnetization protocol is not effective, resulting in large areas of the sample with Type II vertex configurations whose net moments are aligned in the same direction [see magnetic configuration for $\Phi H = 0°$ in Fig. 3(b)]. Instead, we observe relatively large patches of the ground state at $\Phi H = 33.75°$ or 45° [see magnetic configuration for $\Phi H = 45°$ Fig. 3(c)].

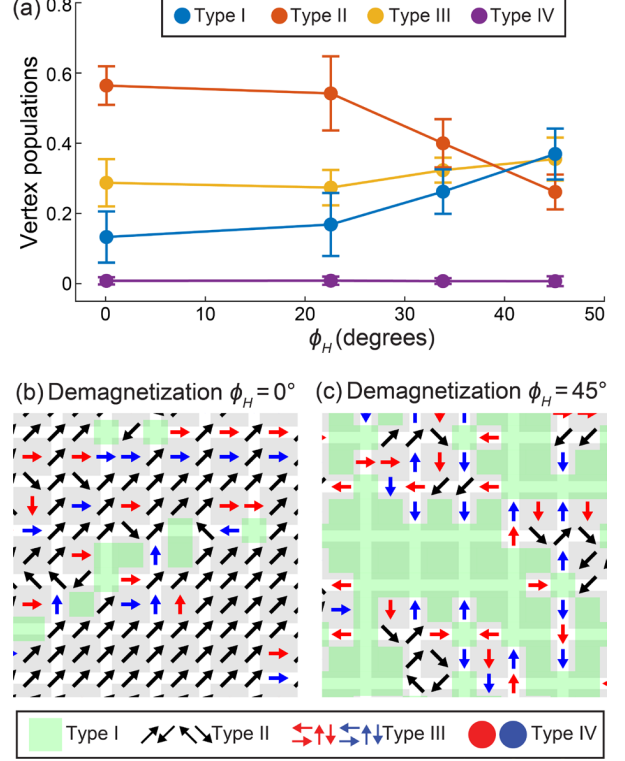

FIG 3. Effectiveness of the demagnetization protocol as a function of the magnetic field $\Phi_H$ angle defined in Fig 1(c). (a) Vertex populations obtained from MFM images after demagnetization with an oscillating bipolar magnetic field $H_A$ applied at different angles $\Phi_H$, while rotating the sample around an axis perpendicular to the substrate plane. Data points and error bars are the mean and standard deviation for 15 nominally identical Pyramid ASIs, respectively. (b-c) Example magnetic configurations for $\Phi H = 0°$ (b) and $\Phi_H = 45°$ (c). Green areas indicate patches of ground state formed by Type I vertex configurations as in the phase diagram in Fig. 2(i). The pyramids appear in (b) and (c) as light gray squares.

By using the magnetic-field demagnetization protocol described above, we have obtained relatively large patches of the ground state for the Pyramid ASI with no $H_Z$ applied. Accessing the other magnetic phases would require a constant out-of-plane magnetic field applied during thermal relaxation or field demagnetization. Thermally annealing the Pyramid ASI with an out-of-plane magnetic field is challenging because nanomagnets that are thick enough to be imaged with MFM require annealing temperatures around 800 K and a 200 nm thick SiN underlayer to prevent oxidation [32]. However, such a thick SiN layer would smooth out the surface altering the surface profile of the supporting pyramid structure. For this reason, we instead demagnetize the Pyramid ASI with an oscillating magnetic field ($H_A$) oriented at $\Phi_H > 45°$ that does not change polarity. The out-of-plane component of $H_A$, which always points in the same direction, forces the macrospins in the tilted nanomagnets at the pyramid vertices into a Type IV configuration. Simultaneously, the sample rotation and the in-plane component of $H_A$ allows the in-plane nanomagnets to reach low-energy configurations (see End Matter, Appendix B for details). With this fixed-polarity demagnetization protocol, we can observe the low-energy magnetic phase for high $H_Z$ [see Fig 2(v) or Fig. 4(a,b)], which features the coarse-grained emergent square ice. In this phase, the effective spins are defined as the net-moments of mixed vertices in a Type III configuration; if a mixed vertex is not in a Type III configuration, it does not contribute a spin to the emergent square ice.

*Contact author: l.berchialla@gmail.com
*Contact author: laura.heyderman@psi.ch

To probe the thermal properties of this emergent square ice, we perform Monte Carlo simulations of the Pyramid ASI and calculate, the heat capacity $c_V$ and the coarse-grained vertex populations [Fig. 4(c)]. We find that the increase in the Type I(coarse-grained) vertex population coincides with the peak in $c_V$, indicating that the ordering of the emergent coarse-grained square ice reflects the onset of long-range order in the Pyramid ASI. This is quite remarkable as, by applying a high $H_Z$, we have engineered a lattice in which a simplified square ice description accurately captures the emergent physics on a larger length scale than that of the underlying nanomagnet array.

We estimate the effective temperature of the experimentally obtained states by finding the best match between the measured coarse-grained vertex populations to those in the simulations. As shown in Fig. 4(c), the experimental data are consistent with a temperature slightly above the ordering transition, with a partially ordered configuration having small domains of Type I(coarse-grained) vertices and short head-to-tail loops of effective macrospins, as seen in the experiment in Fig. 4(b).

## V. CONCLUSIONS

We have introduced a new 3D ASI composed of physically separated, single-domain nanomagnets, which interact solely through dipolar coupling that can be straightforwardly imaged using magnetic force microscopy. The Pyramid ASI design enables selective switching of tilted nanomagnets with an out-of-plane magnetic field and supports multiple magnetic states, including an emergent artificial square ice, where vertex-level degrees of freedom act as effective macrospins representing the first experimental realization of coarse-graining in ASI.

*Contact author: l.berchialla@gmail.com
*Contact author: laura.heyderman@psi.ch

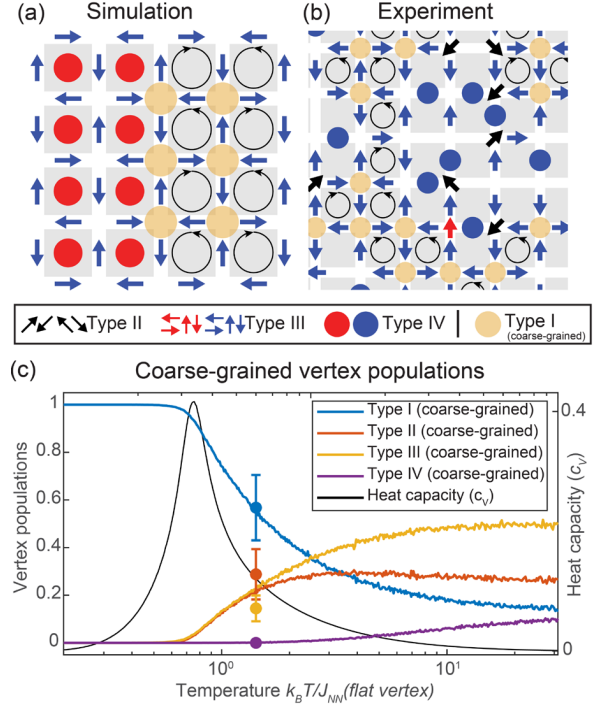

FIG 4. Coarse-grained emergent artificial square ice. (a) Simulated ground state magnetic configuration for high $H_Z$. In the right half, the emergent coarse-grained artificial square ice is highlighted, with head-to-tail loops of emergent macrospins indicated with black circular arrows and emergent Type I(coarse-grained) vertex configurations indicated with yellow circles. (b) Experimental magnetic configuration following demagnetization with the fixed-polarity alternating magnetic field while rotating the sample around an axis perpendicular to the sample plane. For both panels (a) and (b), the symbols are explained in the key and Type I vertex configurations are not displayed to emphasize the emergent square ice ordering. (c) Vertex populations of the coarse-grained square ice as a function of temperature. The solid lines are average values determined from 1000 Monte-Carlo simulations, with the black line corresponding to the heat capacity $c_V$ of the Pyramid ASI. Circular data points and error bars represent the mean and standard deviation of the coarse-grained vertex populations of 15 Pyramid ASIs following demagnetization. The fit of these data points to the simulated lines indicates an effective temperature slightly above the ordering transition.

Moreover, the fabrication process allows for local tuning of the pyramid geometry by altering the pyramid angle and the shape of the base even within the same ASI, opening possibilities to introduce novel forms of vertex or geometrical frustration and can be replicated on different lattices, such as kagome or triangular lattices. For any pyramid lattice geometry, one can then control the appearance of specific phases, magnetic charge ordering [33], as well as magnetic charge propagation with an out-of-plane magnetic field, which is not possible in 2D systems. In addition, with 3D nanomagnet geometries, there is more freedom to create novel magnonic and logic devices based on dipolar coupled nanomagnets that have up to now only been realized in 2D [34,35]. Altogether, this work establishes a versatile platform for investigating emergent behavior and complex magnetic phases in 3D ASIs.

## ACKNOWLEDGMENTS

This work was funded by the Swiss National Science Foundation (project no. 200020_200332). Part of this work was performed on the Merlin6 High Performance Computing Cluster at the Paul Scherrer Institute, Villigen. We thank the staff of the cleanroom facilities at the Park Innovaare (PiA) Cleanroom for Optics and innovation (PICO) for their technical support.

## DATA AVAILABILITY

The data that support the findings of this article will be made openly available on a Zenodo repository. The data will be provided to editors and referees upon request.

[1] A. Fernández-Pacheco, R. Streubel, O. Fruchart, R. Hertel, P. Fischer, and R. P. Cowburn, Three-dimensional nanomagnetism, Nat. Commun. **8**, 15756 (2017).

[2] G. Gubbiotti et al., 2025 roadmap on 3D nanomagnetism, J. Phys.: Condens. Matter **37**, 143502 (2025).

[3] C. Donnelly, M. Guizar-Sicairos, V. Scagnoli, S. Gliga, M. Holler, J. Raabe, and L. J. Heyderman, Three-dimensional magnetization structures revealed with X-ray vector nanotomography, Nature **547**, 328 (2017).

[4] D. Sanz-Hernández et al., Artificial Double-Helix for Geometrical Control of Magnetic Chirality, ACS Nano **14**, 8084 (2020).

[5] A. Hierro-Rodriguez, C. Quirós, A. Sorrentino, L. M. Alvarez-Prado, J. I. Martín, J. M. Alameda, S. McVitie, E. Pereiro, M. Vélez, and S. Ferrer, Revealing 3D magnetization of thin films with soft X-ray tomography: magnetic singularities and topological charges, Nat. Commun. **11**, 6382 (2020).

[6] N. Kent et al., Creation and observation of Hopfions in magnetic multilayer systems, Nat. Commun. **12**, 1562 (2021).

[7] L. Berchialla, G. M. Macauley, and L. J. Heyderman, Focus on three-dimensional artificial spin ice, Appl. Phys. Lett. **125**, 220501 (2024).

[8] A. Farhan, M. Saccone, C. F. Petersen, S. Dhuey, K. Hofhuis, R. Mansell, R. V. Chopdekar, A. Scholl, T. Lippert, and S. van Dijken, Geometrical Frustration and Planar Triangular Antiferromagnetism in Quasi-Three-Dimensional Artificial Spin Architecture, Phys. Rev. Lett. **125**, 267203 (2020).

[9] G. Möller and R. Moessner, Artificial Square Ice and Related Dipolar Nanoarrays, Phys. Rev. Lett. **96**, 237202 (2006).

[10] L. A. S. Mól, W. A. Moura-Melo, and A. R. Pereira, Conditions for free magnetic monopoles in nanoscale square arrays of dipolar spin ice, Phys. Rev. B **82**, 5 (2010).

[11] G.-W. Chern, C. Reichhardt, and C. Nisoli, Realizing three-dimensional artificial spin ice by stacking planar nano-arrays, Appl. Phys. Lett. **104**, 013101 (2014).

[12] Y. Perrin, B. Canals, and N. Rougemaille, Extensive degeneracy, Coulomb phase and magnetic monopoles in artificial square ice, Nature **540**, 7633 (2016).

[13] A. Farhan et al., Emergent magnetic monopole dynamics in macroscopically degenerate artificial spin ice, Sci. Adv. **5**, 2 (2019).

[14] J. de Rojas, D. Atkinson, and A. O. Adeyeye, Tuning magnon spectra via interlayer coupling in pseudo-3D nanostructured artificial spin ice arrays, Appl. Phys. Lett. **123**, 232407 (2023).

[15] J. de Rojas, D. Atkinson, and A. O. Adeyeye, Tailoring magnon modes by extending square, kagome, and trigonal spin ice lattices vertically via interlayer coupling of trilayer nanomagnets, J. Phys.: Condens. Matter **36**, 415805 (2024).

[16] T. Dion et al., Ultrastrong magnon-magnon coupling and chiral spin-texture control in a dipolar 3D multilayered artificial spin-vortex ice, Nat. Commun. **15**, 4077 (2024).

[17] C. Donnelly et al., Element-Specific X-Ray Phase Tomography of 3D Structures at the Nanoscale, Phys. Rev. Lett. **114**, 115501 (2015).

[18] G. Williams et al., Two-photon lithography for 3D magnetic nanostructure fabrication, Nano Res. **11**, 845 (2018).

[19] A. May, M. Hunt, A. Van Den Berg, A. Hejazi, and S. Ladak, Realisation of a frustrated 3D magnetic nanowire lattice, Commun. Phys. **2**, 13 (2019).

*Contact author: l.berchialla@gmail.com
*Contact author: laura.heyderman@psi.ch

[20] A. May, M. Saccone, A. van den Berg, J. Askey, M. Hunt, and S. Ladak, Magnetic charge propagation upon a 3D artificial spin-ice, Nat. Commun. **12**, 3217 (2021).

[21] P. Pip et al., X-ray imaging of the magnetic configuration of a three-dimensional artificial spin ice building block, APL Materials **10**, 101101 (2022).

[22] M. Saccone, A. Van den Berg, E. Harding, S. Singh, S. R. Giblin, F. Flicker, and S. Ladak, Exploring the phase diagram of 3D artificial spin-ice, Commun. Phys. **6**, 217 (2023).

[23] E. Harding et al., Imaging the magnetic nanowire cross section and magnetic ordering within a suspended 3D artificial spin-ice, APL Materials **12**, 021116 (2024).

[24] X. Zhang, I.-A. Chioar, G. Fitez, A. Hurben, M. Saccone, N. S. Bingham, J. Ramberger, C. Leighton, C. Nisoli, and P. Schiffer, Artificial Magnetic Tripod Ice, Phys. Rev. Lett. **131**, 126701 (2023).

[25] S. Ladak, D. E. Read, G. K. Perkins, L. F. Cohen, and W. R. Branford, Direct observation of magnetic monopole defects in an artificial spin-ice system, Nat. Phys. **6**, 359 (2010).

[26] J. Askey, M. O. Hunt, L. Payne, A. van den Berg, I. Pitsios, A. Hejazi, W. Langbein, and S. Ladak, Direct visualization of domain wall pinning in sub-100 nm 3D magnetic nanowires with cross-sectional curvature, Nanoscale **16**, 17793 (2024).

[27] R. F. Wang et al., Artificial 'spin ice' in a geometrically frustrated lattice of nanoscale ferromagnetic islands, Nature **439**, 303 (2006).

[28] See Supplemental Material at [URL will be inserted by publisher] for additional information and figures., (n.d.).

[29] E. O. Wollan and W. C. Koehler, Neutron Diffraction Study of the Magnetic Properties of the Series of Perovskite-Type Compounds $[(1-x)La, xCa]MnO_3$, Phys. Rev. **100**, 545 (1955).

[30] T. Olsen, Antiferromagnetism in two-dimensional materials: progress and computational challenges, 2D Mater. **11**, 033005 (2024).

[31] F. Ott, T. Maurer, G. Chaboussant, Y. Soumare, J.-Y. Piquemal, and G. Viau, Effects of the shape of elongated magnetic particles on the coercive field, J. Appl. Phys. **105**, 013915 (2009).

[32] X. Zhang, Y. Lao, J. Sklenar, N. S. Bingham, J. T. Batley, J. D. Watts, C. Nisoli, C. Leighton, and P. Schiffer, Understanding thermal annealing of artificial spin ice, APL Materials **7**, 111112 (2019).

[33] R. Cheenikundil and R. Hertel, Switchable magnetic frustration in buckyball nanoarchitectures, Appl. Phys. Lett. **118**, 212403 (2021).

[34] A. Imre, G. Csaba, L. Ji, A. Orlov, G. H. Bernstein, and W. Porod, Majority Logic Gate for Magnetic Quantum-Dot Cellular Automata, Science **311**, 205 (2006).

[35] G. Alatteili, A. Roxburgh, and E. Iacocca, Ferromagnetic resonance in three-dimensional tilted-square artificial spin ices, Phys. Rev. B **110**, 144406 (2024).

*Contact author: l.berchialla@gmail.com
*Contact author: laura.heyderman@psi.ch

## END MATTER

### A. Sample Fabrication

Pyramid ASIs are fabricated on $10 \times 10$ mm$^2$ <100> silicon (Si) substrates. The first step consists of creating the pyramid profile in the silicon substrate. Afterwards the nanomagnets are patterned using electron-beam lithography, thermal evaporation and liftoff.

The pyramid profile is defined by the controlled etching of the silicon substrate under a 300 nm-thick PMMA (polymethyl methacrylate) resist mask as illustrated in Fig. 5(a). The PMMA mask is patterned using a Vistec EBPG 5000PlusES electron beam writer at 100 keV accelerating voltage and developed in a 1:3 mixture of methyl isobutyl ketone and isopropanol, before being rinsed with isopropanol, and spin-dried. Subsequently, the silicon substrate is etched in a Oxford PlasmaLab 100 Inductively Coupled Plasma Reactive Ion Etcher (ICP-RIE) for 7 minutes with an RF forward power of 30 W and the following process gas flows: 7 sccm of CHF$_3$, 30 sccm of SF$_6$, 2 sccm of Ar and 2 sccm of O$_2$.

The morphology of the resulting silicon pyramids is significantly influenced by the design of the etch mask. Intuitively, a circular mask should produce a conical structure. However, due to the so-called loading effect, we obtain square pyramids This effect dictates that larger unmasked areas are etched faster than smaller ones due to the higher availability of fresh etching ions and reduced byproduct accumulation.

With circular masks on a square lattice [Fig. 5(b)], there is a larger distance between masks along the diagonal directions than between the masks along the orthogonal ([10] or [01]) directions. Therefore, we obtain square pyramids with the edges of the square, and thus the lateral faces, aligned along the diagonal ([11] and [-11]) directions.

However, for our design, the long axes of the nanomagnets need to be aligned along the orthogonal ([10] or [01]) directions. It is therefore necessary for the pyramid faces to be parallel to these directions. If the faces were aligned along the diagonal ([11] and [-11]) directions, the nanomagnets would be patterned across the edges of the pyramids, resulting in bent or deformed nanomagnets that are no suitable for an ASI. Therefore, to obtain the right orientation of the pyramids we stretched the circular masks in order to fill the gaps in the diagonal directions, resulting in star shaped masks, as shown in Fig. 5(c). With this modified star-shaped mask design, we obtained pyramids with the faces parallel to the [10] or [01] directions.

*Contact author: l.berchialla@gmail.com
*Contact author: laura.heyderman@psi.ch

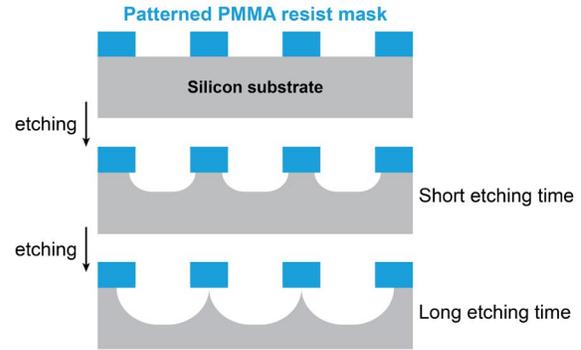

(a) Underetching creates pyramid profile

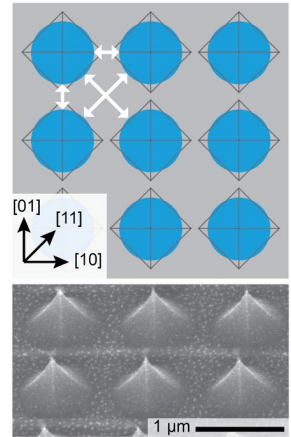

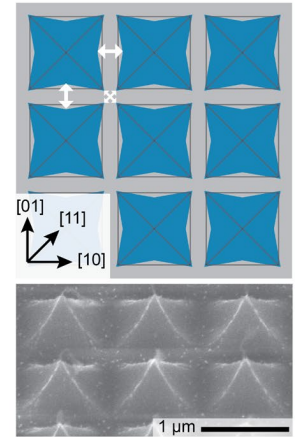

(b) Circular mask     (c) Star mask

FIG 5. (a) Isotropic etching of a silicon substrate through a PMMA resist mask (shown in blue). A short etching time gives trenches in the exposed areas with some underetching (middle schematic). With a longer etching time, the large underetching leads to the creation of pyramid profiles (lower schematic). (b) Circular PMMA etch masks (in blue) give pyramids (outlined in black) with faces parallel to the diagonal ([11] and [-11]) directions. (c) Star-shaped PMMA etch masks (in dark blue) give pyramids with the edges of the square base, and therefore the faces, parallel to the orthogonal ([10] or [01]) directions. The white double-headed arrows indicate the differences in mask separations that give rise to different etch rates, which is responsible for defining the pyramid orientation as explained in the End Matter text.

To fabricate the nanomagnets, the sample is subsequently coated with a 900 nm-thick PMMA resist, which is thick enough to fully cover the pyramids. The resist is then patterned into stadium shaped nanomagnets with lateral dimensions of 450 nm by 120 nm using the electron beam writer. Aside from the necessity to ensure an accurate alignment between the pyramids and the nanomagnets by implementing alignment markers, patterning the nanomagnets with electron-beam lithography does not present any significant challenges, as the depth of focus of the electron beam exceeds the vertical extent of the pyramid profiles. The sample is then developed in a 1:3 mixture of methyl isobutyl ketone and isopropanol, rinsed with isopropanol, and spin-dried. A 20 nm-thick film of Permalloy ($Ni_{80}Fe_{20}$) is then deposited by thermal evaporation at a base pressure of $1 \times 10^{-6}$ mbar and capped, without breaking vacuum, with a 2 nm aluminum layer to prevent oxidation. An ultrasound-assisted lift-off process in acetone was used to remove the undeveloped PMMA resist with the unwanted magnetic material, leaving only the patterned nanomagnets on the sample.

## B. Magnetic-field demagnetization protocols

Pyramid ASI samples are demagnetized using an oscillating magnetic field whose amplitude is continuously reduced, starting from 100 mT in steps of 0.1 mT every 10 seconds. The applied magnetic field $H_A$ follows a sinusoidal profile with a frequency of 2.3 Hz. In order to demagnetize all nanomagnets, the sample is rotated about an axis perpendicular to the substrate surface [Fig. 6(a)] with a frequency at least 100× higher than the frequency of the oscillating magnetic field.

In the demagnetization protocol, illustrated in Fig. 6(b), both the in-plane (xy) and out-of-pane (z) components of $H_A$ oscillate between positive and negative values. In addition, the rotation of the sample adds another sinusoidal oscillation to the in-plane component of the magnetic field affecting each nanomagnet. Therefore, the average magnetic field applied to the sample during a period is zero.

In the fixed-polarity demagnetization protocol, illustrated in Fig. 6(c), both the in-plane (xy) and out-of-pane (z) components of $H_A$ oscillate between zero and a positive value. In this case the rotation of the sample ensures that the in-plane magnetic field affecting each nanomagnet oscillates between positive and negative values. Therefore, the average in-plane magnetic field affecting each nanomagnet during a period is zero. However, the rotation of the sample does not influence the out-of-pane component of the magnetic field affecting each nanomagnet. Therefore, in this unipolar demagnetization protocol there is a non-zero average out-of-pane magnetic field affecting each nanomagnet during each period.

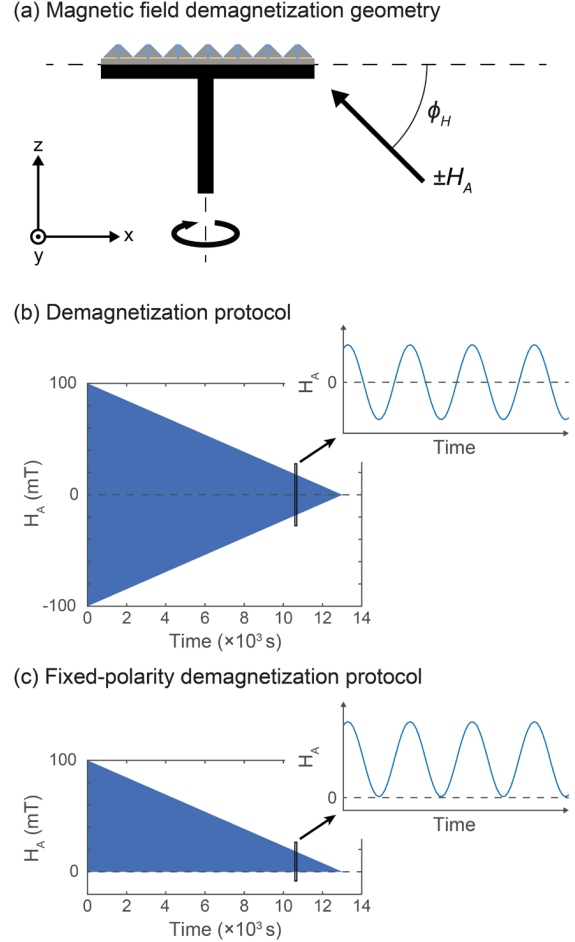

FIG 6. (a) Schematic of the demagnetization setup. The sample is mounted on a holder that rotates about a central axis perpendicular to the substrate surface. The magnetic field $H_A$ is applied at an angle $\Phi_H$ with respect to the substrate surface. (b) For the bipolar demagnetization protocol, the applied magnetic field $H_A$ oscillates between positive and negative values, while for the unipolar demagnetization protocol (c) the applied magnetic field oscillates between zero and positive values. For both cases $H_A$ follows a sinusoidal profile at 2.3 Hz with an amplitude that decreases in steps of 0.1 mT every 10 s.

*Contact author: l.berchialla@gmail.com
*Contact author: laura.heyderman@psi.ch

# Supplemental Material:

# Magnetic-Field Control of Emergent Order in a

# 3D Dipolar Pyramid Artificial Spin Ice

Luca Berchialla,[1,2] Gavin M. Macauley,[1,2,†] Flavien Museur,[1,2] Anja Weber[1,2,‡] and Laura J. Heyderman[1,2]

[1] Laboratory for Mesoscopic Systems, Department of Materials, ETH Zurich, 8093 Zurich, Switzerland

[2] PSI Center for Neutron and Muon Sciences, 5232 Villigen PSI, Switzerland

[†] Present address: Department of Physics, Princeton University, Princeton, NJ 08540 USA

[‡] Present address: PSI Center for Life Sciences, 5232 Villigen PSI, Switzerland

## S1. MONTE CARLO SIMULATIONS

To elucidate the phase diagram of the Pyramid artificial spin ice, Monte Carlo simulations were performed using the Metropolis-Hastings algorithm with a Hamiltonian consisting of two energy terms; the first term is associated with the dipolar interaction between nanomagnets, which are approximated as point dipoles, and the second term corresponds to the Zeeman energy associated with an external magnetic field $\boldsymbol{H_{ext}}$. The Hamiltonian is given by:

$$H = \frac{\mu_0 (M_s V)^2}{4\pi a^3} \sum_{i \neq j}^{N} \left[ \frac{\boldsymbol{s_i} \cdot \boldsymbol{s_j}}{r_{ij}^3} - \frac{3(\boldsymbol{s_i} \cdot r_{ij})(\boldsymbol{s_j} \cdot r_{ij})}{r_{ij}^5} \right] - \mu_0 M_s V \sum_i^N \boldsymbol{s_i} \cdot \boldsymbol{H_{ext}}$$

where, $\boldsymbol{s_i}$ is a unit vector representing the orientation of the magnetic moment at site $i$, located at the position $r_i$. The vector between spin $i$ and $j$ is defined as $r_{ij} \equiv r_j - r_i$. The parameters $M_s$ and $V$ are the saturation magnetization and volume of each nanomagnet, respectively, while $\mu_0$ is the vacuum permeability. The lattice constant $a$, which sets the typical spin–spin distance, is fixed to 1. For convenience, we adopt reduced units $M_s V = 4\pi$, and rescale the temperature and the external magnetic field by dividing by the dipolar energy between the nearest-neighbor spins in a flat vertex $J_{NN}$(flat vertex), which is shown in Fig. S1.

Simulations were carried out on a lattice of 10×10 unit cells, giving a total of 800 spins, with no boundary conditions. The heat capacity, vertex populations and coarse-grained vertex populations, were averaged across 1000 individual simulations. The magnetic configurations in the phase diagram of Fig. 2 correspond to the lowest-energy states obtained via simulated annealing, on cooling from high to low temperature, for each combination of pyramid face angle $\theta_P$ and out-of-plane magnetic field $H_Z$.

We define low, medium and high $H_Z$ according to its effect on the vertex configurations of pyramid vertices in the lowest-energy state reached during simulated annealing. Specifically, Hz is classified as low when pyramid vertices adopt Type I configurations, medium when they adopt Type III configurations and high Hz when they adopt Type IV configurations. The dependence of the pyramid-vertex populations on $H_Z$ for different $\theta_P$ is shown in Fig. S2.

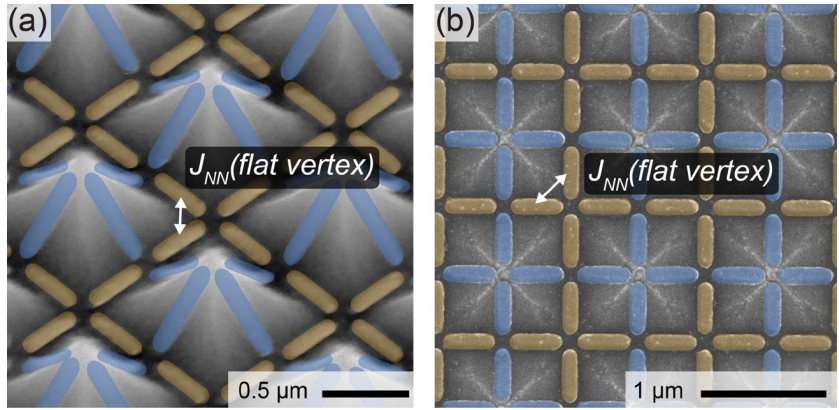

FIG. S1. Scanning electron micrographs of a Pyramid ASI viewed from (a) an oblique angle and (b) from above. Tilted nanomagnets on the lateral faces of the pyramids and in-plane nanomagnets arranged between the pyramids are colored in blue and yellow, respectively. The interaction between nearest-neighbor nanomagnets in a flat vertex $J_{NN}$(flat vertex) is highlighted in both panels with a white arrow.

# Pyramid vertex populations as function of $H_z$ for dfferent $\theta_P$

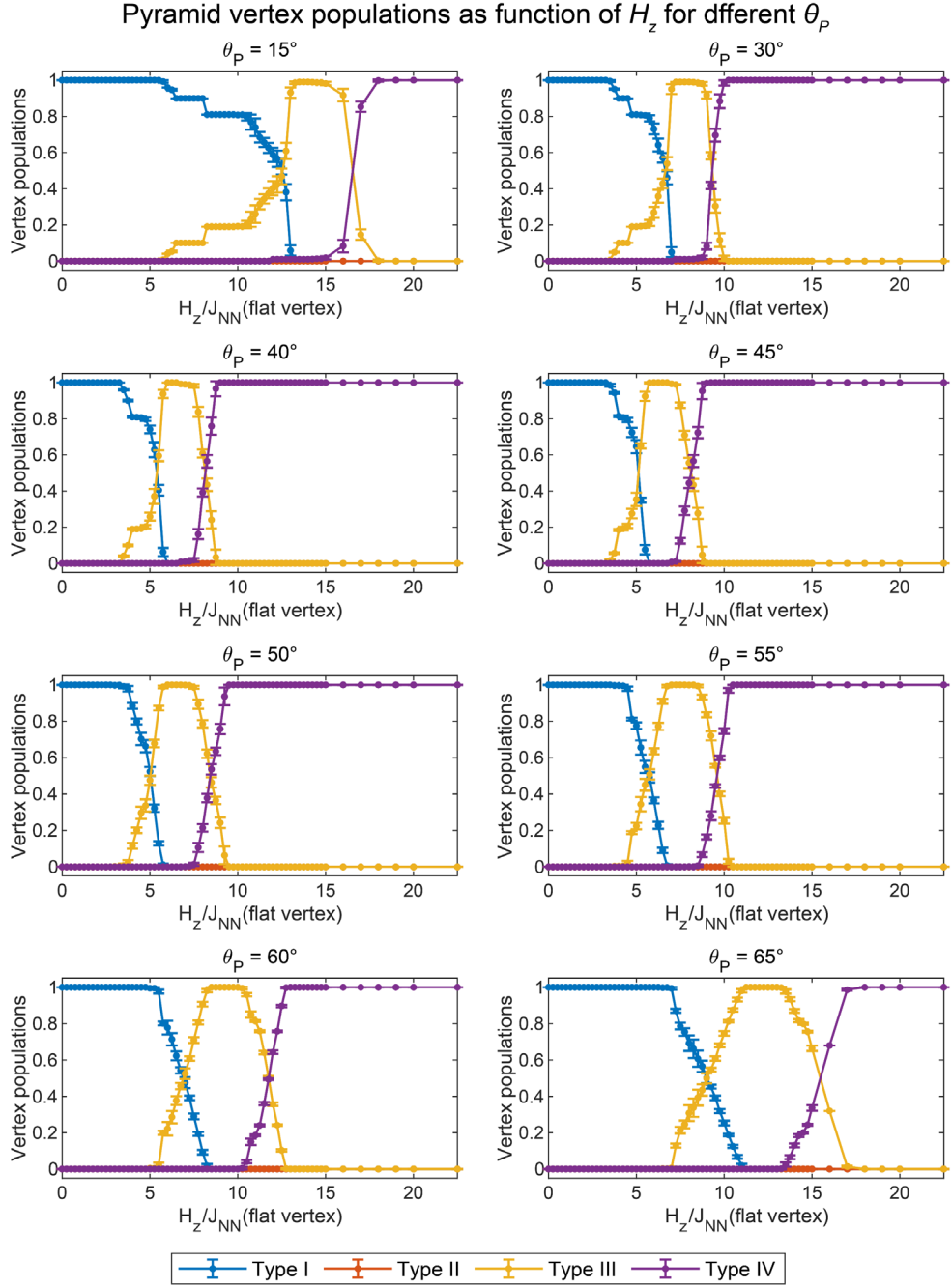

FIG. S2. Vertex populations on pyramid vertices as a function of out-of-plane magnetic field $H_Z$ for pyramid face angles $\theta_P$ = 15°, 30°, 40°, 45°, 50°, 55°, 60°, and 65°. Data points and error bars are the mean and the standard deviation of the vertex populations on pyramid vertices in the lowest-energy states, obtained from 10 Monte-Carlo simulations.

## S2. ENERGY OF VERTEX CONFIGURATIONS AS A FUNCTION OF $\theta_P$

In this section, we present the energy of the different vertex configurations of the macrospins associated with the four magnets at a vertex as a function of the pyramid base angle $\theta_p$, calculated using the point-dipole approximation [Figs. S3(a)-S3(c)]. All possible vertex configurations are illustrated in Fig. S3(d).

For flat vertices, where all four nanomagnets lie in-plane, the energies associated the four vertex types (Type I–IV) remain constant as $\theta_p$ changes [Fig. S3(a)]. This is expected since the geometry of these vertices does not depend on the pyramid shape.

For pyramid vertices, which consist entirely of tilted nanomagnets located on the pyramid faces, the spread of the energies of the vertex configurations increases with pyramid angle $\theta_p$ [Fig. S3(b)], as the dipolar interactions between the pyramid magnets become stronger at larger angles. However, the energy hierarchy remains unchanged with the energy associated with the vertex types $E_{Type\ I} < E_{Type\ II} < E_{Type\ III} < E_{Type\ IV}$, so that a Type I vertex configuration is always the lowest-energy configuration, and a Type IV vertex configuration always has the highest energy.

For mixed vertices, which include two in-plane and two tilted nanomagnets, the energy hierarchy changes with $\theta_p$ [Fig. S3(c)]. At small angles, Type I vertex configurations (characteristic of the ground state in 2D square ice) are energetically favored. As $\theta_p$ increases, the interaction between the two tilted nanomagnets across a vertex strengthens. At $\theta_p \sim 50°$, this interaction becomes dominant, and Type II vertex configurations become the lowest in energy. Furthermore, the absence of four-fold rotational symmetry of mixed vertices due to the presence of both in-plane and tilted nanomagnets lifts the degeneracy of the Type III configurations. We therefore distinguish between Type IIIa vertex configurations, where the macrospins of the two tilted nanomagnets are aligned head-to-tail across the vertex, and Type IIIb vertex configurations, where these macrospins are aligned head-to-head.

As $\theta_p$ increases, the stronger interaction between tilted nanomagnets lowers the energy of the Type IIIa configurations relative to the energy of the Type IIIb configurations.

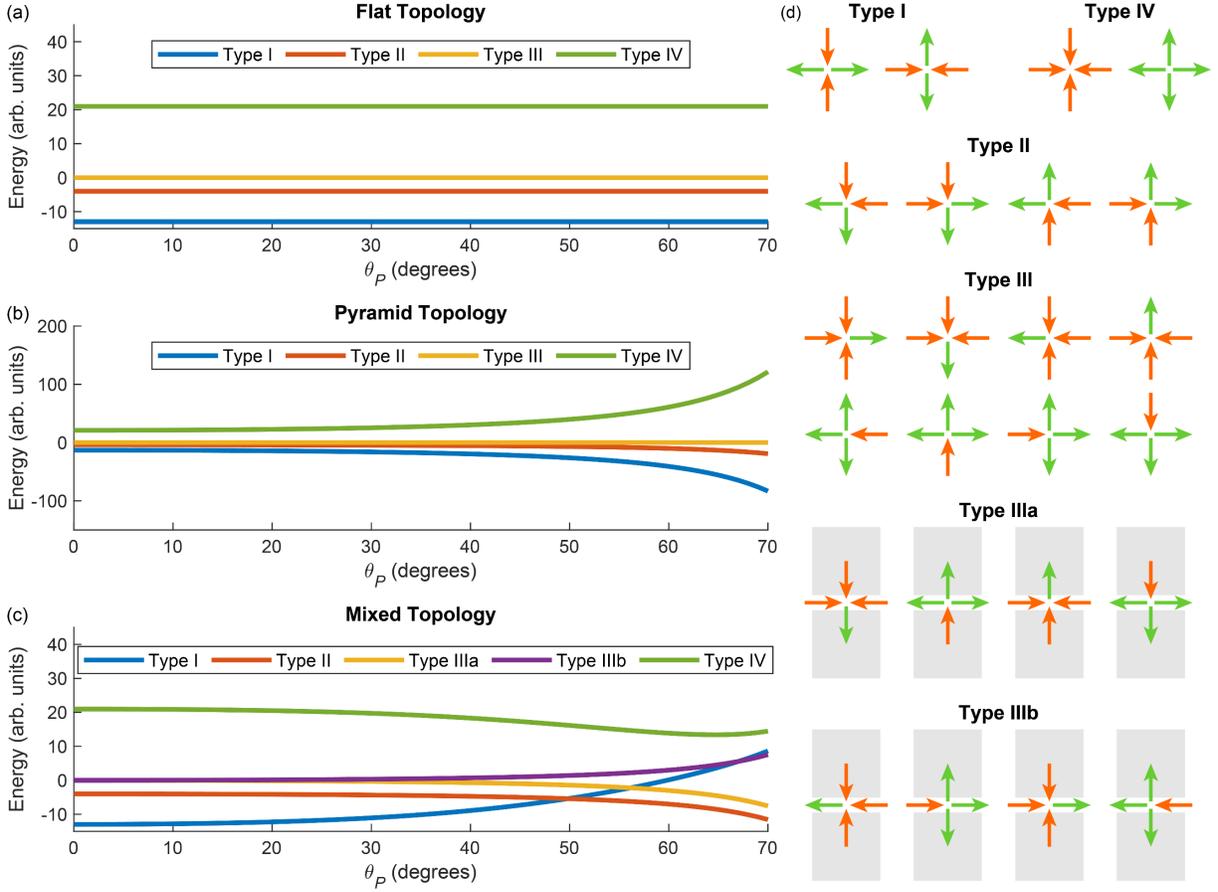

FIG. S3. (a–c) Energy of the vertex configurations as a function of the pyramid angle θ$_p$, calculated using the point-dipole approximation. (a) Flat vertices: the energy of the vertices remains constant with θ$_p$, because their geometry is not affected by the change in angle. (b) Pyramid vertices: the spread of the energies of the vertex configurations increases with θ$_p$, but the energy hierarchy remains unchanged. (c) Mixed vertices: the energy hierarchy changes with θ$_p$; at θ$_p \sim 50°$, Type II vertex configurations become lower in energy than Type I vertex configurations. In addition, the absence of four-fold rotational symmetry of the mixed vertices lifts the degeneracy of Type III configurations. In Type IIIa vertex configurations the macrospins of the two tilted nanomagnets are aligned head-to-tail across the vertex, while in Type IIIb vertex configurations the same macrospins are aligned head-to-head.
(d) Schematic representation of the vertex configurations that arise in the Pyramid artificial spin ice: Type I, II, III and IV vertex configurations, can appear at all vertices. The distinct Type IIIa and Type IIIb vertex configurations appear at mixed vertices.

## S3. COERCIVE FIELD OF NANOMAGNETS WITH DIFFERENT LENGTHS

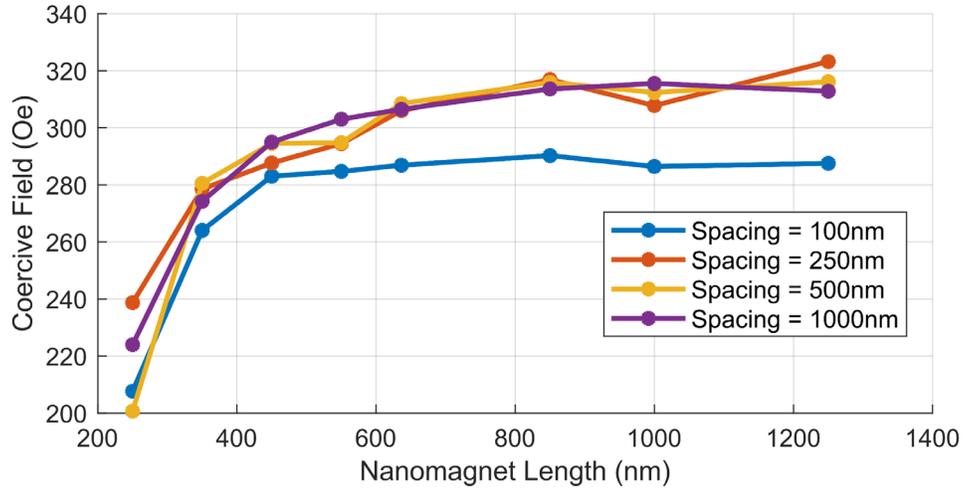

FIG. S4. Coercive field of the stadium shaped nanomagnets determined from hysteresis loops measured along the easy axis with magneto-optical Kerr effect magnetometry as a function of the length of the nanomagnets and the edge-to-edge spacing between them. The nanomagnets are 120 nm wide, as for the nanomagnets in the Pyramid artificial spin ice. The spacing refers to horizontal and vertical distance between the edges of nanomagnets. From these measurements, we estimate that the coercive field of tilted nanomagnets (307 Oe), which are ~636 nm long, to be ~5% higher than the coercive field of the in-plane nanomagnets (292 Oe), which are 450 nm long.

## S4. MAGNETIC FORCE MICROSCOPY

Magnetic force microscopy imaging was carried out using an Asylum Research Jupiter XR atomic force microscope with Bruker MESP-V2-LM tips. Images were acquired in the central part of a 40 µm × 40 µm Pyramid ASI over a 15 µm × 15 µm area with of 512 × 512 pixels, corresponding to a pixel size of approximately 30 nm. Because of the 3D topography of the Pyramid artificial spin ice, a slow scan rate of 0.15 Hz per line was used to ensure accurate surface tracking. The lift height during the second pass was set as low as possible, while avoiding the tip striking the sample, which results in artefacts in the magnetic contrast. Typically, this resulted in lift heights between 30 and 60 nm depending on the condition of the tip.

The vertex populations are extracted from all vertices that are fully present in the image. Therefore, in an image of 15 µm × 15 µm, only an area of approximately 14 µm × 14 µm is considered.

An example of an MFM image is shown in Fig. S4. The same image is shown in Fig. S5, with the edges of the nanomagnets (in white) and the regions used for determining the magnetic configurations (indicated with red frames) overlaid.

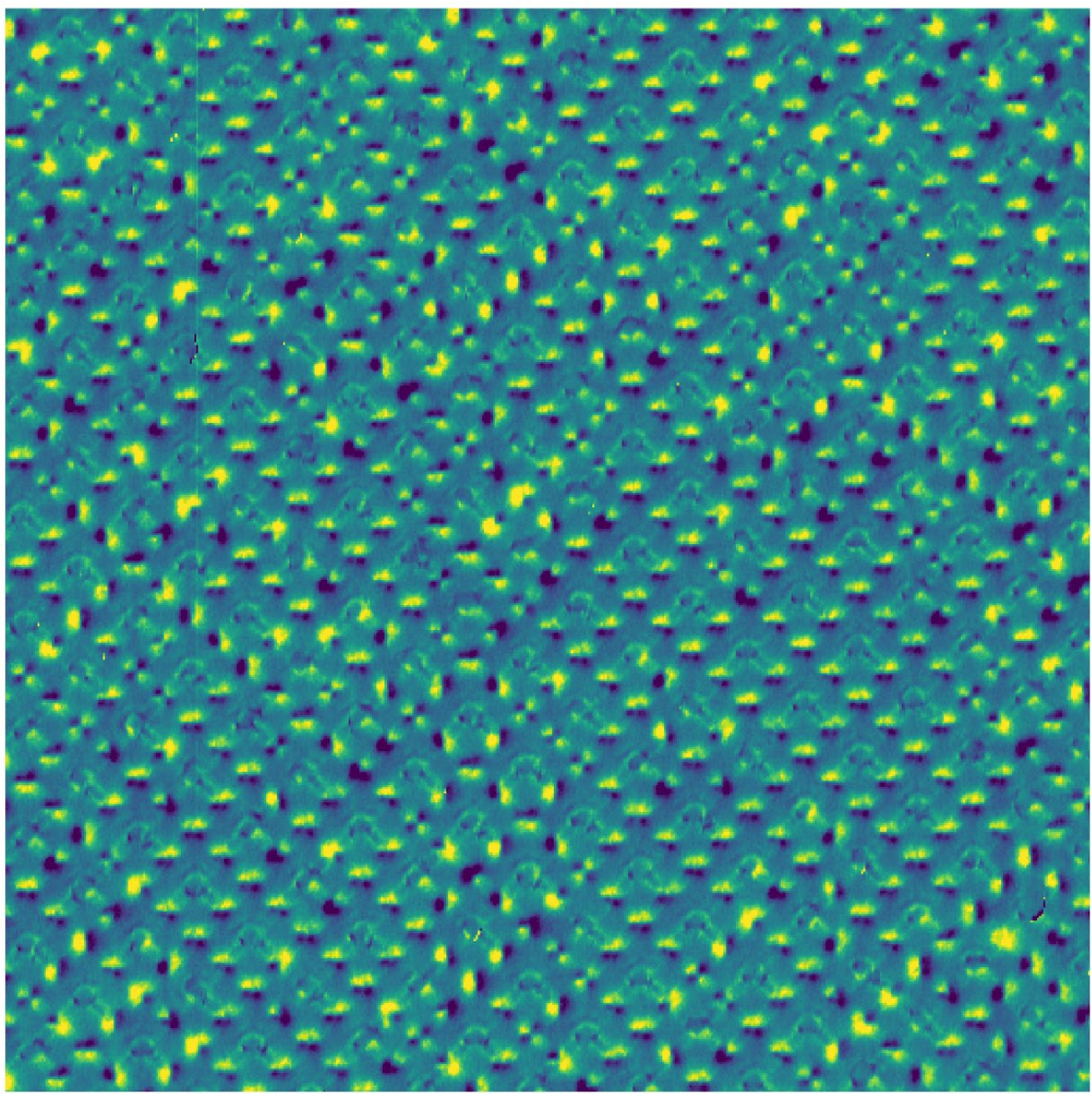

FIG. S5. Example of magnetic contrast obtained using magnetic force microscopy on the Pyramid ASI. The image was acquired over a 15 μm × 15 μm area with 512 × 512 pixels.

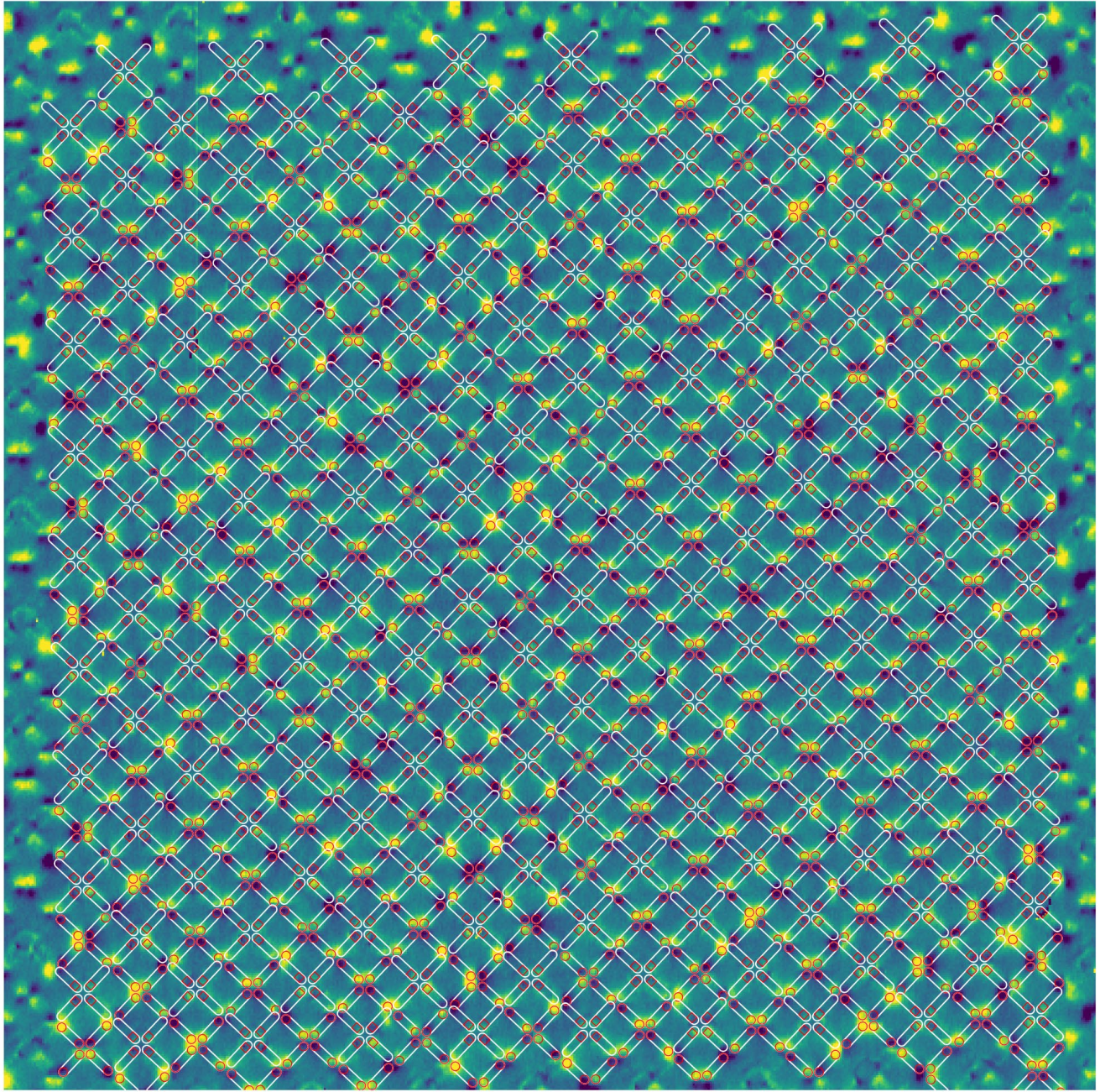

FIG. S6. The same example as in Fig. S5 of magnetic contrast obtained from magnetic force microscopy on the Pyramid ASI, now with the edge of the nanomagnets highlighted in white. In addition, the areas in which the signal is calculated in order to extract the magnetic configuration is shown with red frames; squares for the ends of the nanomagnets at the Pyramid vertices and circles for the ends of all other nanomagnets. The image was acquired over a 15 μm × 15 μm area with 512 × 512 pixels.

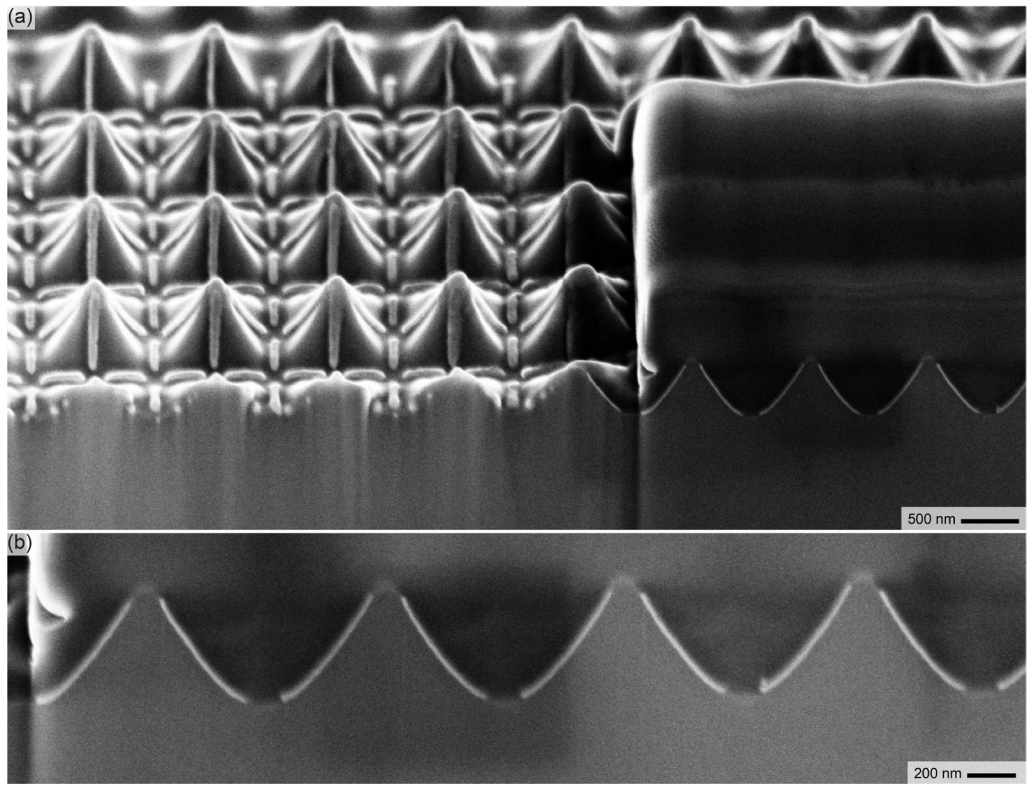

FIG. S7. (a) Scanning electron micrograph of the Pyramid artificial spin ice showing both an oblique overview and a focused ion beam (FIB) cross-section cut. (b) Close-up of the FIB cross-section reveals the pyramid profile etched into the silicon substrate. A slight curvature of the pyramid faces is visible, which leads to minor variations in the thickness of the deposited nanomagnets across each face. The elongated shape of the nanomagnets ensures robust Ising-like behavior associated with the shape anisotropy, making the system resilient to small structural deviations. Therefore, the small curvature of the nanomagnets and non-uniformity in the thickness has a negligible impact on the magnetic properties of the system.